\documentclass{article}
\usepackage[super,compress]{cite}
\usepackage{url, hyperref,amsmath}
\usepackage{amssymb}
\begin{document}

\markboth{Imran Parvez Khan}
{Black Hole partition function and Airy equation}

%\catchline{}{}{}{}{}
%
%%%%%%%%%%%%%%%%%%%%%%%%%%%%%%%%%%%%%%%%%%%%%%%%%%%%%%%%%%%%%%%%%%%%

\title{BLACK HOLE PARTITION FUNCTION AND AIRY EQUATION\footnote{Presented at the 57th Course of the Erice International School of Subnuclear Physics, “In search for the unexpected”, June 2019}}

\author{IMRAN PARVEZ KHAN\footnote{Department of Mathematics, COMSATS University Islamabad,
Islamabad, Pakistan. \newline
Email: imran.parvez@comsats.edu.pk}}

%\address{Department of Mathematics, COMSATS University Islamabad\\
%Islamabad, Pakistan\\
%imran.parvez@comsats.edu.pk}

%\author{SECOND AUTHOR}

%\address{Group, Laboratory, Address\\
%City, State ZIP/Zone, Country\\
%second\_author@group.com}

\maketitle

%\begin{history}
%\received{Day Month Year}
%\revised{Day Month Year}
%\end{history}

\begin{abstract}
We have used the Topological String Theory partition function in the scaling limit, to relate the Black Hole partition function and Topological String Theory partition function.
\end{abstract}

%\keywords{partition function; scaling limit; non-holomorphic terms}

%\ccode{PACS numbers: 04.70.Dy, 04A25}

%\tableofcontents

\section{Introduction}
In this article, we will first introduce the Topological String Theory partition function and add non-holomorphic corrections to it. Then we will employ a scaling limit which will convert the expression for partition function into an Airy equation. We will close the article by expressing Black Hole partition function in terms of a function that is solution of the Airy equation.
\section{Topological String Theory partion function}
The topological string partition function is given by a 					perturbative expansion in the topological string coupling $\lambda$, 			summing the free energies $\mathcal{F}^{(g)}$ over the world sheet 			genera $g$ \cite{Alim:2015qma}
	$$\mathcal{Z}_{top} = \exp \sum \lambda^{2g-2} \mathcal{F}^{(g)} .			$$ 
The perturbative free energies $\mathcal{F}^{(g)}$ satisfy following recursive 			DEs. 
$$ \partial_{i} \mathcal{F}^{(g)} = \frac{1}{2} \left( C_{\overline{i}}^{jk} D_j D_k \mathcal{F}^{(g-1)} + \sum\limits_{r=1}^{g-1} D_j \mathcal{F}^r D_k \mathcal{F}^{(g-r)} \right) $$ 
That were solved using Feynman diagrams \cite{Bershadsky:1993cx}
	but the number of Feynman diagrams grow fast at higher genus and so become unyielding.
This issue was resolved \cite{Yamaguchi:2004bt} due to provision of a polynomial structure of the higher 			genus free energies in finitely many generators.
It is useful to introduce the following propogators $S, S^i , S^{ij} ,$
$$ S \in \Gamma(\mathcal{M} , \mathcal{L}^{-2}) , \quad S^i \in \Gamma (\mathcal{M}, \mathcal{L}^{-2} \otimes T\mathcal{M}), \quad S^{ij} \in \Gamma (\mathcal{M} , \mathcal{L}^{-2} \otimes Sym^2 (T\mathcal{M})) .$$

\subsection{Non-holomorphic corrections}
In order to study Black Holes in string theory, a guiding 				principle is that string theory dualities should remain preserved \cite{Mohaupt:2008gt}.
	 Let $\mathcal{M}$ be the moduli space of a CY threefold Y which is a special $K\ddot{a}hler$ Manifold. The moduli space contains information about coupling constants, among other objects.
	%\item In order to take into account the non-holomorphic corrections, we minimise the free energy w.r.t non-holomorphic section of the slice of moduli space $$\frac{\partial\mathcal{F}^g }{\partial S^{zz} }= 0 .$$
	  The free energy $\mathcal{F}$, in general, is a polynomial in terms of holomorphic and non-holomorphic generators. In some cases, it is possible to consider only the contribution from holomorphic terms to $\mathcal{F}$. But in  case of higher genus mirror symmetry \cite{2014arXiv1409.4105K}, the free energy $\mathcal{F}$ depends not only on holomorphic generators but also on non-holomorphic generators\cite{Alim:2007qj}. Also, in order to preserve the dualities, it is important consider the non-holomorphic dependence of the coupling constants on the moduli.
	In particular we consider the non-holomorphic generator	$S^{zz}$. This and other generators are also called propogators in order to make connection with Feynman diagrams explicit. So from the expression for $\mathcal{F}$, we pick only the terms of the form $f(z) (S^{zz})^{3g-3} $, where $f(z)$ is     a rational function of the modulus $z$.
	 Those terms turn out to be of the form \cite{Bershadsky:1993cx} $f(z)=a_{g} C_{zzz}^{2g-2} ,\quad a_{g} \in \mathbb{Q} . C_{zzz}$ are structure constants that are normalised B-model Yukawa couplings \cite{2014arXiv1409.4105K}.

\subsection{Scaling limit}
The idea of scaling limit is similar to the very important notion of holomorphic limit. Though the following expression contains both holomorphic and non-holomorphic terms, the holomorphic limit allows us to obtain a holomorphic section from the total free energy.
Consider the total free energy$$\mathcal{F} = \sum \lambda^{2g-2} \mathcal{F}^{(g)} .$$
	 We would like to select the terms $a_g C_{zzz}^{2g-2} 					(S^{zz})^{3g-3}$ from $\mathcal{F}$. 
	 To this end, we rescale \cite{Alim:2015qma} the generators $S^{zz}$ and $\lambda$ such that only such terms survive.
\section{Airy equation}
The equation for the partition function $\mathcal{Z}_{top, s} = \exp \mathcal{F}_s$ in the scaling limit becomes $$\left( \left( \frac{1}{3\lambda_s^2} \frac{\partial}{\partial \left( \frac{1}{3\lambda_s^2} \right)} \right)^2  - \left( \left(\frac{1}{3\lambda_s^2} \right)^2 +\frac{1}{9} \right) \right) \lambda_{s} \text{e}^{\frac{1}{3\lambda_s^2}} \mathcal{Z}_{top,s} = 0 ,
	$$ where subscript $s$ denotes scaling limit.
	 Above equation has two singularities. But evidently $\lambda_s = \infty$ is a regular singularity so perturbative expansion is valid for strong coupling. 
	 This can then be further reduced to an Airy equation, via a simple change of variables \cite{Alim:2015qma} $$({\partial_z}^2 - z ) v(z) = 0 .$$
Finally the topological string partition function can be put as $$ \mathcal{Z}_{top,s} = 2^{\frac{1}{3}} \text{e}^{-{\frac{1}{3\lambda^2_s}}} \lambda_s^{-{\frac{1}{3}}} v \left( \left( \frac{1}{2\lambda_s^2} \right)^\frac{2}{3} \right) .$$
	 The function $v$ is a solution of Airy equation. That is a classic DE with two linearly independent solutions. As already pointed out, both solutions can be expanded around strong coupling.

\subsection{Black Hole Partition Function}
This is well known \cite{Mohaupt:2008gt, Ooguri:2004zv} that black hole partition function $Z_{BH}$ is related to the partition function of topological string $$Z_{BH} \approx | \mathcal{Z}_{top} |^2.$$
	 In the special case of scaling limit, this leads to $$Z_{BH} \approx \left| 2^{\frac{1}{3}} \text{e}^{-{\frac{1}{3\lambda^2_s}}} \lambda_s^{-{\frac{1}{3}}} v \left( \left( \frac{1}{2\lambda_s^2} \right)^\frac{2}{3} \right) \right|^{2} .$$

\section{Conclusion and Outlook}
In this article, we have expressed the Black Hole partition function in terms of a function that satisfies Airy equation. 
	 It was made possible due to scaling limit approach to Topological String Theory partition function.
	 Further implications and importance of this result remain to be seen. For instance, the Topological String Theory partition function has a mathematical interpretation as a quantisation of the third cohomology of a CY threefold. What does this mean for $Z_{BH}$?
	 Since Black Hole physics is relatively better understood, such results aid in understanding of the Topological String Theory, despite the fact that we have made use of Topological String Theory to obtain an expression for the Black Hole partition function.

\bibliographystyle{siam}
\bibliography{IPWRA19IPK-ws-ijmpd}

\end{document}